# Different origins or different evolutions?

# Decoding the spectral diversity among C-type asteroids

P. Vernazza[1], J. Castillo-Rogez[2], P. Beck[3], J. Emery[4], R. Brunetto[5], M. Delbo[6], M. Marsset[1], F. Marchis[7], O. Groussin[1], B. Zanda[8,10], P. Lamy[1], L. Jorda[1], O. Mousis[1], A. Delsanti[1], Z. Djouadi[5], Z. Dionnet[5], F. Borondics[9], B. Carry[6]

[1]Aix Marseille Univ, CNRS, LAM, Laboratoire d'Astrophysique de Marseille, Marseille, France

[2]Jet Propulsion Laboratory, California Institute of Technology, 4800 Oak Grove Drive, Pasadena, CA 91109, United States

[3]UJF-Grenoble 1, CNRS-INSU, Institut de Planétologie et d'Astrophysique de Grenoble (IPAG), UMR 5274, Grenoble F-38041, France

[4]Department of Earth & Planetary Sciences and Planetary Geosciences Institute, University of Tennessee, Knoxville, TN 37996-1410, United States

[5]Institut d'Astrophysique Spatiale, CNRS, UMR-8617, Université Paris-Sud, bâtiment 121, F-91405 Orsay Cedex, France

[6]Laboratoire Lagrange, UNS-CNRS, Observatoire de la Cote d'Azur, Boulevard de l'Observatoire-CS 34229, 06304 Nice Cedex 4, France

[7]Carl Sagan Center at the SETI Institute, Mountain View, CA 94043, USA

[8]Institut de Mineralogie, de Physique des Materiaux, et de Cosmochimie (IMPMC), Sorbonne Universites, Museum National d'Histoire Naturelle, UPMC Universite Paris 06, UMR CNRS 7590, IRD UMR 206, 61 rue Buffon, F-75005 Paris, France

[9]SMIS Beamline, Soleil Synchrotron, BP48, L'Orme des Merisiers, 91192 Gif sur Yvette Cedex, France

[10]IMCCE, Observatoire de Paris, 77 avenue Denfert-Rochereau, 75014 Paris Cedex, France

Send correspondence to:

Pierre Vernazza

Laboratoire d'Astrophysique de Marseille

38 rue Frederic Joliot-Curie, 13388 Marseille cedex 13, France

phone : +33 4 91 05 59 11

email : pierre.vernazza@lam.fr

**Abstract**

Anhydrous pyroxene-rich interplanetary dust particles (IDPs) have been proposed as surface analogs for about two thirds of all C-complex asteroids. This suggestion appears, however, inconsistent with the presence of hydrated silicates on the surfaces of some of these asteroids including Ceres. Here we report the presence of enstatite (pyroxene) on the surface of two C-type asteroids (Ceres and Eugenia) based on their spectral properties in the mid-infrared range. The presence of this component is particularly unexpected in the case of Ceres as most thermal evolution models predict a surface consisting of hydrated compounds only. The most plausible scenario is that Ceres' surface has been partially contaminated by exogenous enstatite-rich material, possibly coming from the Beagle asteroid family. This scenario questions a similar origin for Ceres and the remaining C-types, and it possibly supports recent results obtained by the Dawn mission (NASA) that Ceres may have formed in the very outer solar system. Concerning the smaller D~200km C-types such as Eugenia, both their derived surface composition (enstatite and amorphous silicates) and low density (<1.5 g/cm$^3$) suggest that these bodies accreted from the same building blocks, namely chondritic porous, pyroxene-rich IDPs and volatiles (mostly water ice), and that a significant volume fraction of these bodies has remained unaffected by hydrothermal activity likely implying a late accretion. In addition, their current heliocentric distance may best explain the presence or absence of water ice at their surfaces. Finally, we raise the possibility that CI chondrites, Tagish Lake-like material or hydrated IDPs may be representative samples of the cores of these bodies.

# 1. Introduction

C-complex asteroids (B, C, Cb, Cg, Cgh, Ch; DeMeo et al. 2009), which constitute the main population in the asteroid belt (they represent ~38% of the mass of the belt when Ceres, Pallas, Vesta and Hygeia are excluded; ~66% otherwise; DeMeo & Carry 2013), have been extensively studied via spectroscopic measurements in the visible and near-infrared range (0.4-4 µm; Vilas & Gaffey 1989; Vilas et al. 1993; Hiroi et al. 1993, 1996; Rivkin et al. 2003; Clark et al. 2010; Rivkin 2012; Takir & Emery 2012; Vernazza et al. 2015, 2016) and to a lesser extent in the mid infrared domain (7-13 µm range; Barucci et al. 2002; Milliken & Rivkin, 2009; Vernazza et al. 2015; Marsset et al. 2016). Overall, these observations have provided the following constraints:

a) Ch- and Cgh-type asteroids appear to be the parent bodies of the most common type of hydrated meteorites, namely CM chondrites (Vilas & Gaffey 1989; Vilas et al. 1993; Rivkin 2012, Rivkin et al. 2015; Lantz et al. 2013; Burbine 2014; pp. 365–415; McAdam et al. 2015; Vernazza et al. 2016). Furthermore, it appears – on the basis of a spectral comparison in the visible and near-infrared range - that CM chondrites match both the surface and interior of Ch- and Cgh-type asteroids implying a homogeneous internal structure for these bodies (Vernazza et al. 2016). Finally, Ch- and Cgh-type asteroids represent the vast majority of the C-complex asteroids displaying a sharp spectral feature in the 3 µm region (Takir & Emery 2012).

b) The surface material of B-, C-, Cb- and Cg-type asteroids (BCG asteroids hereafter; ~40% of mass of the belt – see DeMeo & Carry 2013) appears to be mostly absent from meteorite collections (see discussion in Vernazza et al. 2015). Instead, interplanetary dust particles (IDPs) may be samples of these objects (Vernazza et al. 2015; Marsset et al. 2016). Specifically, Vernazza et al. (2015) have opened the possibility that pyroxene-rich chondritic porous IDPs (which in fact contain more than 50% amorphous material; Bradley 1994, Ishii et al. 2008) may be the closest analogs to the surfaces of these objects. However, observations of these bodies in the 3-μm region have shown that not all surfaces are compatible with anhydrous silicates as major surface component (e.g., Rivkin et al. 2003; Takir & Emery 2012; De Sanctis et al. 2015, 2016). Some of these objects (e.g., Ceres) display hydrated silicates at their surfaces (e.g., Rivkin et al. 2003; Takir & Emery 2012) as well as carbonates in the case of Ceres (Rivkin et al. 2006; de Sanctis et al. 2015, 2016), minerals that are clearly inconsistent with chondritic porous (anhydrous) IDPs. Observations of these bodies in the 3 μm region also suggest that their surfaces may not only consist of a refractory phase but they may also include volatiles such as water ice (Campins et al. 2010, Rivkin & Emery 2010, Takir & Emery 2012, Combes et al. 2016a,b). In summary, whereas a new interpretation of the surface composition of BCG-types has emerged (e.g., pyroxene-rich IDPs), it does not appear to be always consistent with the spectral diversity observed within this vast group of bodies.

To make progress in our understanding of the surface composition of BCG-type asteroids, we explore the spectral properties of two of them (1) Ceres and (45) Eugenia in the mid-infrared, over the 5-35 μm range. These objects were chosen to investigate, to some extent, the spectral diversity observed in the 3 μm region among C-type bodies as they represent two of the five '3 μm' classes. Four of these classes were defined by Takir & Emery (2012), whereas the last one comprises objects with featureless spectra in this wavelength range. In recent years, it has been shown that mid-infrared spectroscopy is a powerful tool for constraining the surface composition of asteroids of low thermal inertia (i.e., high surface porosity), as their spectra display distinct emissivity features (e.g., Emery et al. 2006, Vernazza et al. 2011, 2012, 2013, 2015). These features can then be used to determine (A) the relative abundance of the main minerals present on the object and/or (B) the meteoritic/IDP analog. Case (A) can be implemented using a spectral decomposition model (e.g., Vernazza et al. 2012, 2015; Marsset et al. 2016) whereas case (B) requires using transmission infrared spectroscopic measurements of IDPs and meteorites (e.g., Sandford & Walker 1985, Izawa et al. 2010, Vernazza et al. 2012, Beck et al. 2014, Merouane et al. 2014). Importantly, sampling the full 7-25 µm range is necessary in order to properly assess the surface composition of an asteroid on the basis of mid-infared spectroscopic observations, provided that its spectrum displays distinct emissivity features (Vernazza et al. 2012). Limiting the compositional analysis to the 7-12 µm range allows detecting crystalline silicates, such as enstatite and olivine, but it prevents the unambiguous detection of amorphous or hydrated silicates. For these latter components, observations at longer wavelengths (12-25 µm range) are required.

## 2. Observations and thermal modeling

2.1 Observations and data reduction

New mid-infrared spectral measurements of (1) Ceres in the ~5–37µm range were obtained using FORCAST, a dual-channel mid-infrared camera and spectrograph, mounted at the Nasmyth focus of the 2.5-m telescope of the Stratospheric Observatory for Infrared Astronomy (SOFIA). The observing run was conducted on June 05$^{th}$, 2015 on a flight originating from Palmdale (California, USA). Detail of the SOFIA observations are summarized in Table 1.

The FORCAST spectrograph with a 2.4 x 194 arcsec slit and a suite of four grisms (4.9 – 8.0 μm, 8.4 – 13.7 μm, 17.6 – 27.7 μm, 28.7 – 37.1 μm) was used in the low-resolution prism mode (R = λ/Δλ ≈ 200) for acquisition of the spectra in the ~5–37μm wavelength range. Images in 4 filters (6.6, 11.1, 19.7 and 34.8 μm) were also secured.

Concerning the data reduction, the SOFIA Science Center provides the processed, flux calibrated Level 3 products for the FORCAST imaging and grism modes. A detailed description of the data reduction steps can be found in the Guest Investigator Handbook for FORCAST Data Products (https://www.sofia.usra.edu/sites/default/files/FORCAST_GI_Handbook_RevB.pdf).

Finally, we complemented our dataset with a Ceres' spectrum retrieved from the ISO archive (http://iso.esac.esa.int/ida/) and with a Spitzer emissivity spectrum for (45) Eugenia taken from Marchis et al (2012).

2.2 Modeling of the thermal emission

We modeled the thermal emission of Ceres in order to derive its infrared emissivity using the Near Earth Asteroid Thermal model (NEATM) developed by Harris (1998), which assumes a spherical body. The NEATM has two free parameters: the radius of the body $r_n$ and the beaming factor η, which accounts for the combined effect of roughness, thermal inertia and geometry (pole orientation, phase angle). We assumed a bond albedo in the visible wavelength range of 0.034 (Li et al. 2016). We computed the temperature distribution at the surface of Ceres for the ISO and SOFIA observing dates (Table 1), from which we derived the Spectral Energy Distribution (SED). Finally, the infrared emissivity was obtained by dividing the observational data by the SED.

The radius $r_n$ controls the intensity of the SED whereas η controls its intensity and shape, and consequently the slope of the infrared emissivity. We adjusted these two parameters in order to obtain a consistent solution between the ISO and SOFIA data in terms of radius, beaming factor and infrared emissivity. For ISO, we obtained a radius of 471 km and a beaming factor of 0.90, and for SOFIA a radius of 468 km and a beaming factor of 0.93. The radii are in excellent agreement with Ceres' mean radius of ~470 km (Park et al. 2016, Russell et al. 2016), whereas the beaming factor close to 1 is consistent with the low thermal inertia of Ceres estimated to be less than 15 $J/m^2/s^{1/2}/K$ by Chamberlain et al. (2011) and in the 5-25 $J/m^2/s^{1/2}/K$ range by Mueller and Lagerros (1998). The ISO and SOFIA data, the modeled SEDs and the corresponding infrared emissivity of Ceres are shown in Fig. 1. We improved the signal-to-noise ratio (SNR) of the emissivity by binning the ISO and SOFIA data. Since the ISO data have a better SNR than the SOFIA data, we used the ISO data whenever available, and the SOFIA data to only fill the ISO spectral gaps. However, we kept the SOFIA data in the 22 − 27 μm range, where the flux is the highest and where it overlaps the ISO data, to provide consistency between the ISO and SOFIA datasets.

## 3. Results

In this section, we present our analysis of the surface composition of (1) Ceres and (45) Eugenia using the two approaches (case A and case B) mentioned in the introduction.

3.1 (1) Ceres

In a first step, we performed a basic reconnaissance of the emissivity features present in the Ceres spectrum and found three features centered at ~7μm, ~10μm and ~22μm that stand out in both the SOFIA and ISO spectra. Whereas the simultaneous presence of the ~10μm and ~22μm features is coherent with the presence of phyllosilicates such as those seen in CI and CM chondrites (e.g., Vernazza et al. 2012, Beck et al. 2014), the ~7μm feature suggests the presence of carbonates at the surface, consistent with previous observations (Rivkin et al. 2006, de Sanctis et al. 2015). Note that ammonium cations within ammoniated phyllosilicates also show a 6.9 μm emissivity feature (Petit et al. 1999). Ceres' ~7μm feature may therefore not only be due to the presence of carbonates at the surface but also to that of ammoniated phyllosilicates (de Sanctis et al. 2015). However, the 3.95 feature observed in Ceres' near-infrared spectrum can only be attributed to carbonates as $NH_4^+$ products do not possess such a spectral band.

In a second step, we performed a detailed comparison of the emissivity spectrum of (1) Ceres with a set of mid-infrared spectroscopic measurements of KB-diluted meteorites (see A.1 and Figure 6). For a quantitative comparison of each meteorite to Ceres, we used the position and width of the diagnostic ~7μm, ~10μm and ~22μm emissivity features as sole criterion rather than the spectral amplitude. We thus allowed the spectral contrast of the meteorite spectrum to vary and minimized the RMS difference between the two spectra.

These comparisons reveal that CI and CM chondrites as well as the ungrouped Tagish Lake meteorite provide the closest - yet unsatisfactory - spectral match to Ceres (Fig. 2). Whereas the two most prominent features in the Ceres spectrum at ~10 μm and ~22 μm are also seen in the meteorite spectra and the weaker feature at ~7 μm is further seen in some of them (especially in the Tagish Lake spectrum), the features in the meteorite spectra (in particular in the 10-μm region) reproduce only partially the spectrum of Ceres. The 10-μm band of all meteorites exhibit a single peak centered at ~10 µm whereas that of Ceres is double-peaked with maxima at ~9.8 and ~10.65 µm. In the 20-µm region, the wide emission band appears narrower in the meteorite spectra than in the Ceres spectrum. Finally, the intensity of the 7-µm band is not well reproduced, implying more carbonates at the surface of Ceres than within the carbonate-rich Tagish Lake meteorite (~12% by volume; Bland et al. 2004) and/or the presence of ammoniated phyllosilicates at the surface of Ceres. These discrepancies observed in the 10- and 20-µm regions open the possibility for at least an additional component at the surface of Ceres that could well be enstatite (a type of pyroxene). Indeed, the spectrum of Ceres in this wavelength region is remarkably similar to that of (10) Hygeia (see A. 2 for discussion and Fig. 7) whose spectrum has been well reproduced assuming enstatite as the dominant surface mineral (Vernazza et al. 2015).

To investigate this possibility, we used a spectral decomposition model as presented in Vernazza et al. (2015) and Marsset et al. (2016). First, the emissivity of enstatite ($Mg_2Si_2O_6$) was calculated for a distribution of hollow spheres with small sizes (<1 µm) compared to wavelength (Min et al. 2003). The optical constants (n and k) of enstatite were retrieved from the Jena database (http://www.astro.uni-jena.de/Laboratory/Database/databases.html). Second, we calculated the best fit to the Ceres spectrum resulting from the linear combination of the emissivity spectra of enstatite and CI/CM chondrites as well as Tagish Lake. Examples of the two best fits are highlighted in Fig. 3. The addition of enstatite not only improves the match in the 10-µm region (i.e., the chi-square is reduced by a factor of two with respect to a best-fit with Tagish Lake only) but it also helps improving the fit in the 20-µm region by better reproducing the width of the emission peak. Additional minerals were tested (crystalline olivine, diposide, amorphous silicates) but none of them reproduced the spectrum of Ceres. In particular, an increasing fraction of crystalline olivine or diopside led to a steady increase of the chi-square value justifying the rejection of these components on Ceres' surface in amounts larger than a few percent. A priori, we cannot exclude the presence of minor amounts (<20%) of amorphous silicates (olivine and pyroxene) at the surface of Ceres as these silicates possess very similar spectral properties to those of hydrated silicates in the 10-µm region. However, their spectral properties in the 20-µm region diverge from those of Ceres with emissivity peaks around ~18 µm. Along these lines, we can also not exclude the presence of minor amounts of iron fragments/particles at the surface, such component being spectrally featureless in the mid-infrared range. Also, the mid-infrared spectral properties of Ceres imply that magnetite can only be a minor (<10%) surface component (Yang & Jewitt 2010) since its emissivity spectrum shows a single peak in the 7-25 µm region centered at ~17 µm, which is at odds with the spectral properties of Ceres. Lastly, we calculated the best fit to Ceres spectrum without enstatite, which was obtained with CI-, CM- or Tagish Lake-like material only (Fig. 2), thus highlighting the unique contribution of enstatite to the improvement of the best fit.

In summary, the surface composition of Ceres appears to be dominated by (i) carbonates, (ii) phyllosilicates (possibly ammoniated), and (iii) enstatite. The presence of anhydrous material (enstatite) on a surface otherwise dominated by products of aqueous alteration (carbonates, phyllosilicates) was unexpected.

3.2 (45) Eugenia

Eugenia is a C-type asteroid with a mean diameter of ~210 km (Marchis et al. 2012). Its mid-infrared spectrum is characterized by two prominent features centered at ~10µm and ~17µm. The 10-µm feature is remarkably similar in shape to that of Ceres whereas the 17-µm feature appears different from any feature observed so far in meteorite spectra. Further comparison of its emissivity spectrum with the spectroscopic measurements of KBr-diluted meteorites reveals no convincing match, suggesting that Eugenia is unsampled by meteorites as already proposed by Vernazza et al. (2015). Importantly, the absence of a ~22 μm feature as observed in the spectra of Ceres and hydrated meteorites implies the absence of phyllosilicates on the surface of Eugenia. This conclusion is well supported by measurements at shorter wavelengths that revealed an absence of hydration band in the 3-μm region (Rivkin et al. 2003).

A spectral decomposition model with the same minerals as in Vernazza et al. (2015) reproduces the spectral behavior of Eugenia with crystalline pyroxene (enstatite; ~10%) and amorphous silicates of olivine-like composition (~90%) only (Fig. 4). Note that the mid-infrared spectrum of amorphous silicates of pyroxene-like composition differs only slightly from the one of olivine-like composition. Therefore, the constraint on the mineralogy of the amorphous component is less compelling than for the crystalline one.

## 4. Discussion

Our analysis of the surface composition of (1) Ceres and (45) Eugenia appears not only coherent with previous findings based on spectroscopic observations in the 3 μm region (e.g., Rivkin et al. 2003, 2006, de Sanctis et 2015, 2016) but it is further complementary. Concerning Ceres, early observations by Rivkin et al. (2003) in the 3-μm region showed that its surface is less hydrated than that of CM-like Ch- and Cgh-types (by ~30%) opening the possibility of the presence of an anhydrous component at its surface. Later on, carbonates were detected by Rivkin et al. (2006) and further confirmed by observations with the VIR spectrometer of the Dawn mission (de Sanctis et al. 2015, 2016), which further unambiguously identified phyllosilicates (de Sanctis et al. 2015). Enstatite, however, has not been previously reported as a main surface component for Ceres. Note that an additional dark and neutral component such as carbon may well be present at the surface (Hendrix et al. 2016) in order to explain the overall low reflectivity of Ceres (Bond Albedo~0.035; Li et al. 2016). Concerning Eugenia, Rivkin et al. (2003) found its surface to be essentially anhydrous. Our spectral analysis not only confirms this early result but also provides a more detailed characterization of the mineralogical composition of its surface (mixture of enstatite and amorphous silicates).

Hereafter, we first discuss the probable exogenous origin of enstatite on the surface of Ceres. We further underline that this 'enstatite anomaly' is also observed in the case of Hygiea, the second largest C-type asteroid. We next attempt to formulate a comprehensive scenario for the formation and evolution of IDP-like D~200km C-type bodies. In particular, we introduce a new hypothesis for the composition of the putative cores of these bodies. In this respect, hydrated extraterrestrial materials such as CI chondrites, the Tagish Lake meteorite, and hydrous IDPs appear as plausible candidates.

4.1 On the origin of the presence of enstatite on the surface of Ceres

4.1.1 Enstatite as an endogenous component

A first explanation for the presence of enstatite on the surface of Ceres is that such material was part of the primordial building blocks of Ceres along with water ice and that a fraction of this material was preserved intact during Ceres' early thermal evolution. Up to now, however, none of the numerical models simulating this evolution (e.g., McCord & Sotin

2005, Castillo-Rogez & McCord 2010, Castillo-Rogez 2011; Neveu and Desch 2015) have produced a compositional stratification that would naturally lead to the present surface being a mixture of hydrous and primordial anhydrous silicates following impact mixing. Most simulations (e.g., McCord & Sotin 2005, Castillo-Rogez & McCord 2010), which assume that Ceres accreted from a mixture of ice and anhydrous material in roughly equal proportions in agreement with the Dawn-derived density estimate (2.16 g/cm$^3$; Park et al. 2016, Russell et al. 2016), predict that the primordial crust would have foundered as a consequence of global melting and impacting, which has been simulated by Formisano et al. (2016). Following these models, the present crust should entirely consist of products of aqueous alteration. Note that one of the models by McCord and Sotin (2005) assumed an intact - 130 km thick - primordial anhydrous outer shell (their case 1) superimposed on an hydrated, phyllosilicate-rich core. This scenario seems unlikely, as it is highly improbable that impacts over the course of solar system history could have mixed the two layers and produced the homogeneity of the hydrated material found on the surface (Ammannito et al. 2016).

Finally, we discuss hereafter two scenarios that seem highly unlikely to explain the coexistence of anhydrous and hydrous silicates at the surface of Ceres:

a) It is unlikely that the anhydrous silicates could originate from the dehydration of the phyllosilicates following micrometeorite bombardment. First, micrometeorite impacts do not produce enough energy to bring the surface temperature to the required >500 deg. C temperature (King et al. 2015). Second, this effect would also apply to many other asteroids, such as Pallas and/or CM-like bodies, yet this is not supported by observations.

b) It also seems unlikely that silicates would be subject to partial aqueous alteration in Ceres' near-surface. This scenario implies a dearth of water. However the abundance of Mg-serpentine (antigorite) (de Sanctis et al. 2015) implies advanced alteration leading to the leaching of iron from the original anhydrous silicates. Furthermore, gravity and topography observations point to an abundance of volatiles at depth (Fu et al. submitted).

4.1.2 Enstatite as an exogenous component

A more plausible explanation for the presence of enstatite on the surface of Ceres is impact implantation of exogenous material. Prior works have demonstrated that impactors can survive during hypervelocity collisions with porous targets and thus remain intact at their surfaces (Daly and Schultz, 2015, 2016; Avdellidou et al. 2016; Turrini et al. 2016). Retention of exogenous material may not only work for porous targets but for compact ones as well. Indeed, the presence of dark material on Vesta's surface (McCord et al., 2012; Reddy et al., 2012; Turrini et al., 2014) has been interpreted as direct evidence for projectile retention. A recent study also raised the possibility that the presence of olivine at several locations on Vesta's surface (Ammannito et al. 2013; Ruesch et al. 2014; Palomba et al. 2015, Poulet et al. 2015) may be best explained as impact contamination (Turrini et al. 2016).

From above, the hypothesis that impact-delivered debris could have been retained on Ceres appears plausible and Ceres' present surface composition may therefore not only reflect its endogenic evolution but also that of exogenous compounds (Daly and Schultz 2015). In this scenario, it remains to be understood whether the exogenic products (20-40% volume fraction following our spectral analysis) essentially originated from a few large impacts or from numerous smaller contributions by IDPs. The first case (a few large impacts) seems unlikely at first glance if one considers that Hygiea, the second largest C-type, presents spectral properties similar to those of Ceres (see A.2 and Fig. 7), implying not only a similar formation and thermal evolution, but also contamination by similar enstatite-rich material. It seems improbable that two of the four largest main belt asteroids were impacted by compositionally similar asteroids, especially if one considers that the two bodies reside in different regions of the asteroid belt (middle belt for Ceres, outer belt for Hygiea). Contamination by IDPs would therefore appear as a more plausible explanation for the presence of enstatite at the surface of Ceres under the assumption that its origin is exogenous. IDP fluxes in this region of the solar system amount to approximately $4 \times 10^{-5}$ particles/m$^2$/s (Grün et al. 2001) implying that a square millimeter is impacted by about $\sim 10^3$ particles over $\sim 10^6$ years. Considering a typical IDP size of 10-100 μm, these high fluxes may effectively be compatible with asteroid surfaces being spectrally (at least in the mid-infrared where small particles such as IDPs dominate the emissivity) a mixture of endogenous and exogenous materials. If this hypothesis is correct, enstatite should be found at the surface of compositionally diverse asteroids located in the vicinity of both Ceres and Hygiea. In addition, a contaminating enstatite-rich dust source should exist in the outer belt.

Based on current knowledge, there is one large outer main belt family whose composition is compatible with pyroxene-rich IDPs, namely the Themis family (Marsset et al. 2016). Within this family, a recent break up (<10 Myrs) has led to the formation of the Beagle family located at semi major axis a = 3.157 AU (Nesvorny et al. 2008). Spitzer Space Telescope observations and numerical modeling have shown that the Beagle family is the most likely source of the prominent α dust band (Nesvorny et al. 2008). Furthermore, metallic asteroids located in the vicinity of Ceres and Hygiea show spectral evidence of anhydrous pyroxene-rich dust at their surface (Hardersen et al. 2005) whereas olivine should be observed in case of differentiation. As in the case of Ceres, we suggest an exogenous origin for this dust.

Importantly, an exogenous origin for enstatite questions a similar origin for Ceres, Hygiea and the remaining C-types such as Eugenia and Themis. As far as Ceres' phyllosilicates are concerned (ammoniated), such different origin is well supported (de Sanctis et al. 2015): Ceres may have formed in the very outer solar system (>10 AU), beyond the ammonia snowline, and may be a close relative of large TNOs such as Orcus. Conversely, enstatite-rich C-types such as Eugenia likely formed in the inner solar system (<10 AU), in agreement with model calculations of the dust composition in a stationary protoplanetary solar disk as a function of heliocentric distance (Gail 2004).

4.2 Constraining the formation and evolution of D<200km IDP-like C-type asteroids

In this section, we apply our results along with those produced by Takir and Emery (2012), Vernazza et al. (2015) and Marsset et al. (2016) to formulate a comprehensive scenario for the formation and evolution of IDP-like D<~200km BCG-type asteroids (B, C, Cb, Cg). Spectral similarity in both the 0.4-2.5 µm and 10-µm regions (Vernazza et al. 2015, Marsset et al. 2016) suggests similar compositions for the surface of these objects and thus opens the possibility that they accreted from the same building blocks. Both the density (in the 0.8-1.5 g/cm$^3$ range; Marchis et al. 2012 and references therein) and the inferred surface composition of these bodies (this work; Vernazza et al. 2015, Marsset et al. 2016) suggest that these building blocks included volatiles (especially water ice) and pyroxene-rich IDP-like material consisting of a minor phase of crystalline pyroxene and a major phase of amorphous silicates. Observations in the 3-µm region (Rivkin et al. 2003; Campins et al. 2010; Rivkin & Emery 2010; Takir & Emery 2012) have further revealed that the surfaces of these objects consist of water ice and/or anhydrous silicates.

Overall, these observations support a scenario where these objects have suffered limited aqueous alteration and have consequently preserved intact at least part of their primordial outer shell. Indeed amorphous material is very metastable in hydrothermal environment and rapidly turns into hydrated silicates even at low temperatures (e.g., Nakamura-Messenger et al. 2011). In addition, the outer shell must have been sufficiently thick to prevent either impact excavation of inner hydrated silicates in the course of the solar system history (as suggested in the case of Saturn's satellite Phoebe; Castillo-Rogez et al. 2012) or collapse of the primordial crust into a subsurface 'ocean' during the early thermal evolution (as suggested in the case of Ceres). In support of our analysis, Beauvalet & Marchis (2014) showed that Eugenia likely possesses a differentiated interior, consisting of a low-porosity ~3 g/cm$^3$ dense core (<15% in volume) and a porous icy shell that is ~70–80 km thick. This internal structure could be consistent with the result of limited early aqueous alteration due to the decay of short lived radionuclides leading to the formation of a small phyllosilicate-rich core (Fig. 5). The fact that a significant volume fraction of these bodies has remained unaffected by hydrothermal activity likely imply a late accretion for these bodies. Note that a similar conclusion had been reached for the Themis parent body (Marsset et al. 2016). The internal structure of Eugenia derived by Beauvalet & Marchis (2014) would naturally explain why the hydrated silicates never reached the surface via impacts and therefore the anhydrous nature of its surface (Fig. 5).

In summary, one cannot exclude that IDP-like BCG-type bodies all formed in the same accretional environment, possibly in the giant planet regions, and were scattered across the solar system following, for example, giant planet migrations, as advocated by dynamical models (e.g., Walsh et al. 2011). Their final heliocentric distance would then determine the ability of these bodies to retain water ice on their surfaces in large amounts and/or for a long period of time, in agreement with theoretical estimates of the stability of water ice at the surface/near surface of main belt asteroids (e.g., Schorghofer 2008). Note that future surveys in the 3 μm region will eventually find a phyllosilicate hydration feature in the spectra of some of these objects whose hydrated interior has been excavated by collisions. This is expected in particular for D<~100km bodies that are believed to be collisional fragments rather than primordial undisrupted bodies (Morbidelli et al. 2009).

4.3 Cores of IDP-like asteroids: CI- or Tagish Lake-like material?

In the previous section, we have focused our attention on the surface composition of a large (D~200km) IDP-like C-type asteroid. So far, spectroscopic surveys in the 3 μm region and in the mid-infrared range of these low albedo C-type bodies ($p_v \leq 0.1$) have mostly focused on the largest objects (D>~100 km), thus mostly on primordial asteroids that did not suffer a catastrophic disruption (Morbidelli et al. 2009). Thus, it is mostly the composition of the primordial outer shell of these bodies that has been properly characterized so far, and the latter appears to be consistent with anhydrous pyroxene-rich IDP-like material (Vernazza et al. 2015; Marsset et al. 2016). We can extend this inference to D>100km P- and D-type asteroids, whose primordial outer shell mainly consists of a mixture of anhydrous olivine and pyroxene-rich IDP-like material (e.g., Vernazza et al. 2012, 2015).

Concerning the internal compositional structure of these bodies, there are preliminary hints that they may possess a differentiated interior (Marchis et al. 2014; Beauvalet & Marchis 2014). This differentiation can be well understood as a natural consequence of their early thermal evolution if one assumes that these bodies accreted from a mixture of ice and dust. In this scenario, the core would likely consists of a phyllosilicate-rich material.

Extraterrestrial material collections contain a small number of phyllosilicate-rich samples, including CI chondrites, CM chondrites, the atypical Tagish Lake meteorite (e.g., Vernazza et al. 2013 and references therein) and hydrated IDPs. Whereas a large number of potential parent bodies have been identified for CM chondrites (e.g., Vernazza et al. 2016), the same cannot be said for CI chondrites (see section 11 in Cloutis et al. 2011), Tagish Lake (Vernazza et al. 2013), and hydrated IDPs. The paucity of CI-like bodies is particularly striking since CI falls amount to only one third of the CM ones. Considering that CM-like asteroids, namely Ch- and Cgh-types, are routinely observed and represent about one third of the C-complex asteroids (Rivkin 2012) and that CIs are significantly more friable than CMs, implying a bias against their preservation during atmospheric entry, one would expect CI-like bodies to be commonly observed as well. Yet, this is not the case. In a similar register, the paucity of Tagish-Lake like bodies has already been noticed by Vernazza et al. (2013).

One potential explanation for this paradox could be that CIs, Tagish Lake and hydrated IDPs are not representative of the original surface material of primordial D>100km asteroids. Instead, these aqueously altered materials may be samples of the cores of IDP-like C, P or D-type (D>100km) asteroids. In fact, the water-rich, chondrule-free CIs could potentially represent an aqueously altered, consolidated version of anhydrous IDP-like dust. In other words, CIs could be representative materials of the cores of IDP-like >100 km large C-, P- and D-type bodies that initially accreted from a mixture of ice and dust and where thermal evolution led to the early formation of a phyllosilicate-rich CI-like core (see 4.2; e.g., Beauvalet & Marchis 2014). For similar reasons, both Tagish Lake and hydrated IDPs could also be samples of the cores of IDP-like C-, P- and D-type bodies.

This conjecture implies that CI-like and Tagish Lake-like asteroids may have to be searched among smaller, D<50 km, asteroids that would be leftover fragments of the catastrophic disruption of D>100 km IDP-like C-, P- and D-type bodies. In this respect, collisional families offer a great opportunity to investigate the spectral properties and thus composition of these cores. The Eurybates family, the only well-identified collisional family among the Jupiter Trojans (P/D types), whose parent body was estimated to be in the D~110-130 km size range (Broz & Rozehnal 2011), is particularly relevant here. Spectroscopic observations of the family members have revealed a great spectral variety (Fornasier et al. 2007; DeLuise et al. 2010), with about half of the family members being C-types and the remaining objects being P-types. Considering that the largest Jupiter Trojans are either P- or D-type asteroids, these observations could suggest that the cores of P-/D-type asteroids may consist of C-type-like material. Since CI chondrites possess mostly C-type-like spectral properties, this opens the possibility that CI chondrites could be samples of the cores of P/D-type asteroids. Note that is seems quite unlikely that space weathering weathering processes could be at the origin of the slope variation within the family (i.e., Cs and Ps) given that the surface age should be roughly the same for all objects. If there was only one C-type, one could argue for a recent resurfacing event but in the present case the similar frequency of Cs and Ps tends to exclude the space-weathering hypothesis.

These preliminary ideas lead to new questions to be addressed in future research:

- Could CI chondrites and Tagish Lake represent samples of the cores of D>100 km IDP-like C-, P- and D-type bodies? Or would hydrated IDPs be more relevant analogs?

- What is the expected compositional difference between the core of an IDP-like C-type (e.g., Eugenia) and the core of an IDP-like D-type (e.g., Hektor) and what are their respective extraterrestrial analogs (if any)?

## 5. Conclusion

Spectroscopic observations in the 5-35 μm range of Ceres and Eugenia enabled a deeper investigation of the compositional diversity among C-type asteroids. Our derived compositions appear not only consistent with previous interpretations derived from spectral analyses performed at shorter wavelengths (in the 3 μm region in particular) but they are also complementary. Both the 'wet' and 'dry' nature of the surfaces of respectively Ceres and Eugenia could be revealed in the mid-infrared range. In addition, our investigation allowed the detection and subsequent characterization of the anhydrous silicate composition present at the surfaces of both objects. Specifically, enstatite is likely present on the surface of Ceres at a level of at least 20vol.% mixed with products of aqueous alteration (Dawn VIR observations could only detect the latter), whereas both enstatite and amorphous silicates are the main surface components of Eugenia. Furthermore, our analysis confirms earlier suggestions that carbonaceous chondrite meteorites are poor analogs for the refractory material present on these objects. Importantly, the presence of enstatite at the surface of Ceres is striking and unexpected. The most plausible explanation for the origin of this component is an exogenous delivery of pyroxene-rich dust possibly coming from the Beagle family. This scenario is compatible with, yet not implies, an outer solar system origin for Ceres, as indicated by the detection of ammoniated phyllosilicates on its surface by the Dawn mission (de Sanctis et al., 2015).

For the smaller D<~200km C-types, we propose that these bodies accreted from the same building blocks, namely chondritic porous, pyroxene-rich (and amorphous-rich) IDPs and volatiles (mostly water ice), and that a significant volume fraction of these bodies has remained unaffected by hydrothermal activity likely implying a late accretion for these bodies. In addition, their current heliocentric distance may best explain the presence or absence of water ice at their surfaces. We also introduce a new hypothesis regarding the composition of the cores of these bodies: hydrated extraterrestrial materials such as CI chondrites, the Tagish Lake meteorite, and hydrous IDPs appear plausible candidates.

Finally, it is important to stress that the future of asteroid science would greatly benefit from a space-based observatory dedicated to the observations of large (D>100 km) and medium-sized (20 km<D<100 km) asteroids in the ~5-25 μm spectral range. Datasets collected with ISO and Spitzer for these asteroids exist only for a few objects (< 30). That type of observation cannot be performed with the James Webb Space Telescope as objects with D>20 km will be too bright for that telescope (Rivkin et al. 2016).

**Acknowledgments**:

We thank the referee for his pertinent and constructive remarks. This work is based on observations made with the NASA/DLR Stratospheric Observatory for Infrared Astronomy (SOFIA). SOFIA is jointly operated by the Universities Space Research Association, Inc. (USRA), under NASA contract NAS2-97001, and the Deutsches SOFIA Institut (DSI) under DLR contract 50 OK 0901 to the University of Stuttgart. We warmly thank Melody Didier for producing figure 5's artwork.

**Appendix**

A.1 Laboratory data for meteorites

Here, we present new mid-infrared laboratory emissivity spectra of 13 carbonaceous chondrite meteorites collected over the 16.6-35 μm range. These new spectra complement those acquired by Beck et al. (2014) over the 2-25 μm range for the same samples. Accordingly, a detailed description of the sample preparation can be found in Beck et al. (2014).

New mid-IR micro-analysis was performed at the SMIS beamline (Dumas et al., 2006) of the synchrotron SOLEIL (France). We used a NicPlan microscope (x32 objective) coupled to a NEXUS 6700 FTIR spectrometer (Thermo Fisher) operating in transmission, in a similar configuration previously used to analyze cometary dust (see Brunetto et al., 2011 for more details). The analyzed spot diameter was 20-30 μm on the samples, so that several positions were analyzed and spectra were averaged to increase the signal to noise ratio. Spectra were collected in the 600–285 cm$^{-1}$ (16.6–35 μm) spectral range, because the KBr matrix is poorly transparent for wavelength larger than ~35 μm. In the 25-35 μm spectral range, a correction was applied to take into account the small KBr absorption based on the knowledge of the KBr optical constants (Handbook of optical constants of solids, edited by E.D. Palik (1985) Academic Press) and the thickness of the pellets, we estimated the contribution of the KBr to the absorption and subtracted it from the measured spectra (error bars are <5% in the 25-30 μm range, and <10% in the 30-35 μm range). The spectra were finally adjusted and combined to the Beck et al. spectra by using the overlapping spectral range, and they are presented in Fig. 6.

A.2 Similarities and differences in composition between Ceres and Hygiea

(10) Hygiea is the only known main belt asteroids that possesses spectral properties nearly similar to (1) Ceres in the 0.4-2.5 µm range, in the 3-µm range, and in the 10-µm and 20-µm regions (see Fig. 7). There are, however, three noticeable spectral differences between the two objects. The first is the absence of a carbonate band around ~3.95 µm in the spectrum of Hygiea. The second is a shift in the 20-µm peak found at Hygiea towards shorter wavelengths (~20 µm) with respect to Ceres (~22 µm). The third is a shallower depth of the ~3.06 µm feature in the spectrum of Hygiea with respect to Ceres. All these differences indicate that the surface of Hygiea contains less hydrated minerals than Ceres; conversely, they suggest that anhydrous silicates are more abundant at the surface of Hygiea than Ceres. The mid-infrared range supports this assumption particularly well since anhydrous silicates such as amorphous and crystalline pyroxene possess emissivity peaks around ~20 µm whereas phyllosilicates peak around ~22 µm.

**References**

Ammannito, E., de Sanctis, M. C., Palomba, E., Longobardo, A., Mittlefehldt, D. W., et al. Olivine in an unexpected location on Vesta's surface. Nature 504, 122-125 (2013).

Ammannito, E., DeSanctis, M. C., Ciarniello, M., Frigeri, A.; Carrozzo, F. G., et al. Distribution of phyllosilicates on the surface of Ceres, Science 353 (2016).

Avdellidou, C., Price, M. C., Delbo, M., Ioannidis, P., Cole, M. J. Survival of the impactor during hypervelocity collisions - I. An analogue for low porosity targets. MNRAS 456, 2957-2965 (2016).

Barucci, M. A., Dotto, E., Brucato, J. R., Müller, T. G., Morris, P., et al. 10 Hygiea: ISO Infrared Observations. Icarus 156, 202-210 (2002).

Beauvalet, L. & Marchis, F. Multiple asteroid systems (45) Eugenia and (87) Sylvia: Sensitivity to external and internal perturbations. Icarus 241, 13-25 (2014).

Beck, P. et al. Transmission infrared spectra (2-25 mum) of carbonaceous chondrites (CI, CM, CV-CK, CR, C2 ungrouped): Mineralogy, water, and asteroidal processes. Icarus, Volume 229, p. 263-277 (2014a).

Bland, P.A., Cressey, G., Menzies, O.N. Modal mineralogy of carbonaceous chondrites by X-ray diffraction and Mössbauer spectroscopy. Meteorit. Planet. Sci. 39, 3–16 (2004).

Bradley, J. P. Chemically Anomalous, Preaccretionally Irradiated Grains in Interplanetary Dust From Comets. Science 265, 925-929 (1994).

Brunetto, R., Borg, J., Dartois, E., Rietmeijer, F. J. M., Grossemy, F., et al. Mid-IR, Far-IR, Raman micro-spectroscopy, and FESEM-EDX study of IDP L2021C5: Clues to its origin. Icarus 212, 896-910 (2011).

Brož, M. & Rozehnal, J. Eurybates - the only asteroid family among Trojans? Monthly Notices of the Royal Astronomical Society 414, 565-574 (2011).

Burbine, T. H. Asteroids. Planets, Asteriods, Comets and The Solar System, Volume 2 of Treatise on Geochemistry (Second Edition). Edited by Andrew M. Davis. Elsevier, 365-415 (2014).

Bus, S. J. Compositional Structure in the Asteroid Belt: Results of a Spectroscopic Survey. Thesis, Massachusetts Inst. Technol. (1999).

Campins, H., Hargrove, K., Pinilla-Alonso, N., Howell, E. S., et al. Water ice and organics on the surface of the asteroid 24 Themis. Nature 464, 1320-1321 (2010).

Castillo-Rogez, J. C. & McCord, T. B. Ceres' evolution and present state constrained by shape data. Icarus 205, 443-459 (2010).

Castillo-Rogez, J. C. Ceres - Neither a porous nor salty ball. Icarus 215, 599-602 (2011).

Castillo-Rogez, J. C., T. V. Johnson, P. C. Thomas, M. Choukroun, D. L. Matson, J. I. Lunine. Geophysical evolution of Saturn's satellite Phoebe, a large satellite planetesimal in the outer solar system. Icarus 219, 86-109 (2012).

Chamberlain, Matthew A., Lovell, Amy J., Sykes, Mark V. Submillimeter photometry and lightcurves of Ceres and other large asteroids. Icarus 202, 487-501 (2009).

Clark, B. E., Ziffer, J., Nesvorny, D., Campins, H., Rivkin, A. S. et al. Spectroscopy of B-type asteroids: Subgroups and meteorite analogs. Journal of Geophysical Research 115, Issue E6, CiteID E06005 (2010).

Cloutis, E. A., Hiroi, T., Gaffey, M. J., Alexander, C. M. O.'D., Mann, P. Spectral reflectance properties of carbonaceous chondrites: 1. CI chondrites. Icarus 212, 180-209 (2011).

Combe, J.-Ph., McCord, T. B., Tosi, F., Raponi, A., De Sanctis, M. C., et al. Detection of H2O-Rich Materials on Ceres by the Dawn Mission. 47th Lunar and Planetary Science Conference, Contribution No. 1903, p.1820 (2016a).

Combe, J.-Ph., McCord, T. B., Tosi, F., Ammannito, E., Carrozzo, F. G., et al. Detection of local $H_2O$ exposed at the surface of Ceres. Science 353, 6p (2016b).

Daly, R. T., Schultz, P. H. Predictions for impactor contamination on Ceres based on hypervelocity impact experiments. Geophysical Research Letters 42, 7890-7898 (2015).

Daly, R. T., Schultz, P. H. Delivering a projectile component to the vestan regolith. Icarus 264, 9-19 (2016).

De Luise, F., Dotto, E., Fornasier, S., Barucci, M. A., Pinilla-Alonso, N., et al. A peculiar family of Jupiter Trojans: The Eurybates. Icarus 209, 586-590 (2010).

DeMeo, F. E., Binzel, R. P., Slivan, S. M., Bus, S. J. An extension of the Bus asteroid taxonomy into the near-infrared. Icarus 202, 160-180 (2009).

DeMeo, F. E.&Carry, B. The taxonomic distribution of asteroids from multi-filter all sky photometric surveys. Icarus 226, 723–741 (2013).

De Sanctis, M. C., Ammannito, E., Raponi, A., Marchi, S., McCord, T. B., et al. Ammoniated phyllosilicates with a likely outer Solar System origin on (1) Ceres. Nature 528, 241-244 (2015).

De Sanctis, M. C., Raponi, A., Ammannito, E., Ciarniello, M., Toplis, M. J., et al. Bright carbonate deposits as evidence of aqueous alteration on (1) Ceres. Nature 536, 54-57 (2016).

Dumas, P., Polack, F., Lagarde, B., Chubar, O., Giorgetta, J. L., Lefrançois, S. Synchrotron infrared microscopy at the French Synchrotron Facility SOLEIL. Infrared Physics & Technology 49, 152-160 (2006).

Emery, J.P., Cruikshank, D.P., van Cleve, J. Thermal emission spectroscopy (5.2–38 μm) of three Trojan asteroids with the Spitzer Space Telescope: Detection of fine-grained silicates. Icarus 182, 496–512 (2006).

Formisano, M., Federico, C., De Angelis, S., De Sanctis, M. C., Magni, G. The stability of the crust of the dwarf planet Ceres. Monthly Notices of the Royal Astronomical Society 463, 520-528.

Fornasier, S., Dotto, E., Hainaut, O.; Marzari, F., Boehnhardt, H., et al. Visible spectroscopic and photometric survey of Jupiter Trojans: Final results on dynamical families. Icarus 190, 622-642 (2007).

Gail, H. P. Radial mixing in protoplanetary accretion disks. IV. Metamorphosis of the silicate dust complex. Astronomy and Astrophysics 413, 571-591 (2004).

Groussin, O., Lamy, P., Fornasier, S., Jorda, L. The properties of asteroid (2867) Steins from Spitzer Space Telescope observations and OSIRIS shape reconstruction. Astronomy & Astrophysics 529, id.A73, 8 pp. (2011).

Grün, E., Baguhl, M., Svedhem, H., Zook, H. A. In situ Measurements of Cosmic Dust. Interplanetary Dust, Edited by E. Grün, B.A.S. Gustafson, S. Dermott, and H. Fechtig. Astronomy and Astrophysics Library. Berlin: Springer, p295 (2001).

Hardersen, P. S., Gaffey, M. J., Abell, P. A. Near-IR spectral evidence for the presence of iron-poor orthopyroxenes on the surfaces of six M-type asteroids. Icarus 175, 141-158 (2005).

Harris, A. W. A Thermal Model for Near-Earth Asteroids. Icarus 131, 291-301 (1998).

Hendrix, A. R., Vilas, F. and Li, J-Y. Ceres: Sulfur deposits and graphitized carbon. GRL 43, (2016).

Hiroi, T., Pieters, C. M., Zolensky, M. E., Lipschutz, M. E. Evidence of thermal metamorphism on the C, G, B, and F asteroids. Science 261, 1016-1018 (1993).

Hiroi, T., Zolensky, M. E., Pieters, C. M., & Lipschutz, M. E. Thermal metamorphism of the C, G, B, and F asteroids seen from the 0.7 μm, 3 μm and UV absorption strengths in comparison with carbonaceous chondrites. Meteor. Planet. Sci. 31, 321 (1996).

Ishii, H. A., Bradley, J. P., Dai, Z. R., et al. Comparison of Comet 81P/Wild 2 Dust with Interplanetary Dust from Comets. Science 319, 447
(2008).

Izawa, M. R. M., King, P. L., Flemming, R. L., Peterson, R. C., McCausland, P. J. A. Mineralogical and spectroscopic investigation of enstatite chondrites by X-ray diffraction and infrared reflectance spectroscopy. Journal of Geophysical Research 115, CiteID E07008 (2010).

King, A. J., Schofield, P. F., Russell, S. S. Thermal Alteration of CI and CM Chondrites: Mineralogical Changes and Metamorphic Temperatures. 78th Annual Meeting of the Meteoritical Society, LPI Contribution No. 1856, p.5212 (2015).

Lantz, C., Clark, B. E., Barucci, M. A., Lauretta, D. S. Evidence for the effects of space weathering spectral signatures on low albedo asteroids. Astronomy & Astrophysics 554, 7 pp. (2013).

Li, J.-Y., Reddy, V., Nathues, A., Le Corre, L., Izawa, M. R. M., et al. Surface Albedo and Spectral Variability of Ceres. The Astrophysical Journal Letters 817, id. L22, 7 pp. (2016).

McAdam, M. M., Sunshine, J. M., Howard, K. T., McCoy, T. M. Aqueous alteration on asteroids: Linking the mineralogy and spectroscopy of CM and CI chondrites. Icarus 245, 320-332 (2015).

Marchis, F., Descamps, P., Baek, M., Harris, A. W., Kaasalainen, M., et al. Main belt binary asteroidal systems with circular mutual orbits. Icarus 196, 97-118 (2008).

Marchis, F., Enriquez, J. E., Emery, J. P., Mueller, M., Baek, M., et al. Multiple asteroid systems: Dimensions and thermal properties from Spitzer Space Telescope and ground-based observations. Icarus 221, 1130-1161 (2012).

Marsset, M., Vernazza, P., Birlan, M., DeMeo, F., Binzel, R. P., et al. Compositional characterisation of the Themis family. A&A 586, 9 pp. (2016).

McCord, T. B. & Sotin, C. Ceres: Evolution and current state. Journal of Geophysical Research 110, Issue E5, CiteID E05009 (2005).

McCord, T. B., Li, J.-Y., Combe, J.-P., McSween, H. Y., Jaumann, R., et al. Dark material on Vesta from the infall of carbonaceous volatile-rich material. Nature 491, 83-86 (2012).

Merouane, S., Djouadi, Z., Le Sergeant d'Hendecourt, L. Relations between Aliphatics and Silicate Components in 12 Stratospheric Particles Deduced from Vibrational Spectroscopy. The Astrophysical Journal 780, 12 pp. (2014).

Milliken, R. E. & Rivkin, A. S. Brucite and carbonate assemblages from altered olivine-rich materials on Ceres. Nature Geoscience 2, 258-261 (2009).

Min, M., Hovenier, J. W., de Koter, A. Shape effects in scattering and absorption by randomly oriented particles small compared to the wavelength. Astronomy and Astrophysics 404, 35-46 (2003).

Morbidelli, A, Bottke, W. F., Nesvorný, D., Levison, H. F. Asteroids were born big. Icarus 204, 558-573 (2009).

Mueller, T. G. & Lagerros, J. S. V. Asteroids as far-infrared photometric standards for ISOPHOT. Astronomy and Astrophysics 338, 340-352 (1998).

Nakamura-Messenger, K., Clemett, S. J., Messenger, S., Keller, L. P. Experimental aqueous alteration of cometary dust. Meteorit. Planet. Sci. 46, 843-856 (2011).

Nesvorný, D., Bottke, W. F., Vokrouhlický, D., Sykes, M., Lien, D. J., Stansberry, J. Origin of the Near-Ecliptic Circumsolar Dust Band. The Astrophysical Journal Letters 679, article id. L143, pp. (2008).

Neveu, M. & Desch, S. J. Geochemistry, thermal evolution, and cryovolcanism on Ceres with a muddy ice mantle. Geophysical Research Letters 42, 197-206 (2015).

Park, R. S., Konopliv, A. S., Bills, B., Castillo-Rogez, J., Asmar, S. W., et al. Gravity Science Investigation of Ceres from Dawn. 47th Lunar and Planetary Science Conference, LPI Contribution No. 1903, p.1781 (2016).

Palik, E. D. Handbook of optical constants of solids. Academic Press Handbook Series, New York: Academic Press, edited by Palik, Edward D (1985).

Palomba, E., Longobardo, A., De Sanctis, M. C., Zinzi, A., Ammannito, E., et al. Detection of new olivine-rich locations on Vesta. Icarus 258, 120-134 (2015).

Petit, S., Righi, D., Madejova, J., Decarreau, A. Interpretation of the infrared $NH_4^+$ spectrum of the $NH_4^+$-clays; application to the evaluation of the layer charge. Clay Minerals 34, 543-549 (1999).

Poulet, F., Ruesch, O., Langevin, Y., Hiesinger, H. Modal mineralogy of the surface of Vesta: Evidence for ubiquitous olivine and identification of meteorite analogue. Icarus 253, 364-377 (2015).

Prettyman, T. H., Yamashita, N., Castillo-Rogez, J. C., Feldman, W. C., Lawrence, D. J., et al. Elemental Composition of Ceres by Dawn's Gamma Ray and Neutron Detector. 47th Lunar and Planetary Science Conference, LPI Contribution No. 1903, p.2228 (2016).

Reddy, V., Le Corre, L., O'Brien, D. P., Nathues, A., Cloutis, E. A., et al. Delivery of dark material to Vesta via carbonaceous chondritic impacts. Icarus 221, 544-559 (2012).

Rivkin, A. S. et al. Hydrogen concentrations on C-class asteroids derived from remote sensing. Meteoritics &Planetary Science 38, 1383-1398 (2003).

Rivkin, A. S., Volquardsen, E. L., Clark, B. E. The surface composition of Ceres: Discovery of carbonates and iron-rich clays. Icarus 185, 563-567 (2006).

Rivkin, A. S. & Emery, J. P. Detection of ice and organics on an asteroidal surface. Nature 464, 1322-1323 (2010).

Rivkin, A. S. The fraction of hydrated C-complex asteroids in the asteroid belt from SDSS data. Icarus 221, 744-752 (2012).

Rivkin, A. S., Thomas, C. A., Howell, E. S., Emery, J. P. The Ch-class Asteroids: Connecting a Visible Taxonomic Class to a 3 μm Band Shape. The Astronomical Journal 150, 14 pp. (2015).

Rivkin, A. S., Marchis, F., Stansberry, J. A., Takir, D., Thomas, C., et al. Asteroids and the James Webb Space Telescope. Publications of the Astronomical Society of Pacific 128, pp. 018003 (2016).

Ruesch, O., Hiesinger, H., De Sanctis, M. C., Ammannito, E., Palomba, E., et al. Detections and geologic context of local enrichments in olivine on Vesta with VIR/Dawn data. Journal of Geophysical Research: Planets 119, 2078-2108 (2014).

Russell, C. T., Raymond, C. A., Ammannito, E., Buczkowski, D. L., De Sanctis, M. C. et al. Dawn arrives at Ceres: Exploration of a small, volatile-rich world. Science 353, 1008-1010 (2016).

Sandford, S. A. & Walker, R. M. Laboratory infrared transmission spectra of individual interplanetary dust particles from 2.5 to 25 μm. Astrophysical Journal 291, 838-851 (1985).

Schorghofer, N. The Lifetime of Ice on Main Belt Asteroids. The Astrophysical Journal 682, 697-705 (2008).

Schmidt, B. E. & Castillo-Rogez, J. C. Water, heat, bombardment: The evolution and current state of (2) Pallas. Icarus 218, 478-488 (2012).

Schmidt, B. E., Hughson, K. G., Chilton, H. T., Scully, J. E. C., Platz, T., et al. Ground Ice on Ceres? 47th Lunar and Planetary Science Conference, LPI Contribution No. 1903, p.2677 (2016).

Sizemore, H. G., Platz, T., Schorghofer, N., Mest, S. C., Crown, D. A., et al. Preliminary Constraints on the Volumetric Concentration of Shallow Ground Ice on Ceres from Geomorphology. 47th Lunar and Planetary Science Conference, LPI Contribution No. 1903, p.1628 (2016).

Takir, D. & Emery, J. P. Outer Main Belt asteroids: Identification and distribution of four 3-mum spectral groups. Icarus 219, 641-654 (2012).

Turrini, D., Combe, J.-P., McCord, T. B., Oklay, N., Vincent, J.-B., et al. The contamination of the surface of Vesta by impacts and the delivery of the dark material. Icarus 240, 86-102 (2014).

Turrini, D., Svetsov, V., Consolmagno, G., Sirono, S., Pirani, S. Olivine on Vesta as exogenous contaminants brought by impacts: Constraints from modeling Vesta's collisional history and from impact simulations. Icarus 280, 328-339 (2016).

Vernazza, P., Lamy, P., Groussin, O., Hiroi, T., Jorda, L. et al. Asteroid (21) Lutetia as a remnant of Earth's precursor planetesimals. Icarus 216, 650-659 (2011).

Vernazza, P. et al. High surface porosity as the origin of emissivity features in asteroid spectra. Icarus 221, 1162–1172 (2012).

Vernazza, P., Fulvio, D., Brunetto, R., Emery, J. P., Dukes, C. A., et al. Paucity of Tagish Lake-like parent bodies in the Asteroid Belt and among Jupiter Trojans. Icarus 225, 517-525 (2013).

Vernazza, P., Marsset, M., Beck, P., Binzel, R. P.; Birlan, M. et al. Interplanetary Dust Particles as Samples of Icy Asteroids. The Astrophysical Journal 806, 10 pp. (2015).

Vernazza, P., Marsset, M., Beck, P., Binzel, R. P., Birlan, M., et al. Compositional Homogeneity of CM Parent Bodies. The Astronomical Journal 152, article id. 54, pp. (2016).

Vilas, F. & Gaffey, M. J. Phyllosilicate absorption features in main-belt and outer-belt asteroid reflectance spectra. Science 246, 790-792 (1989).

Vilas, F., Larson, S. M., Hatch, E. C., Jarvis, K. S. CCD Reflectance Spectra of Selected Asteroids. II. Low-Albedo Asteroid Spectra and Data Extraction Techniques. Icarus 105, 67-78 (1993).

Walsh, K. J., Morbidelli, A., Raymond, S. N., O'Brien, D. P. & Mandell, A. M. A low mass for Mars from Jupiter's early gas-driven migration. Nature 475, 206–209 (2011).

Yang, B. & Jewitt, D. Identification of Magnetite in B-type Asteroids. The Astronomical Journal 140, 692-698 (2010).

**Table 1: Observing parameters of our SOFIA observations**

| Asteroid | Observing date | $r^a$ (AU) | $\Delta^b$ (AU) | $\alpha^c$ (°) |
|---|---|---|---|---|
| 1 Ceres | 05 June 2015 | 2.92 | 2.22 | 16.5 |

[a] r is the heliocentric distance.

[b] Δ is the SOFIA–asteroid distance.

[c] α is the phase angle.

**Figures**

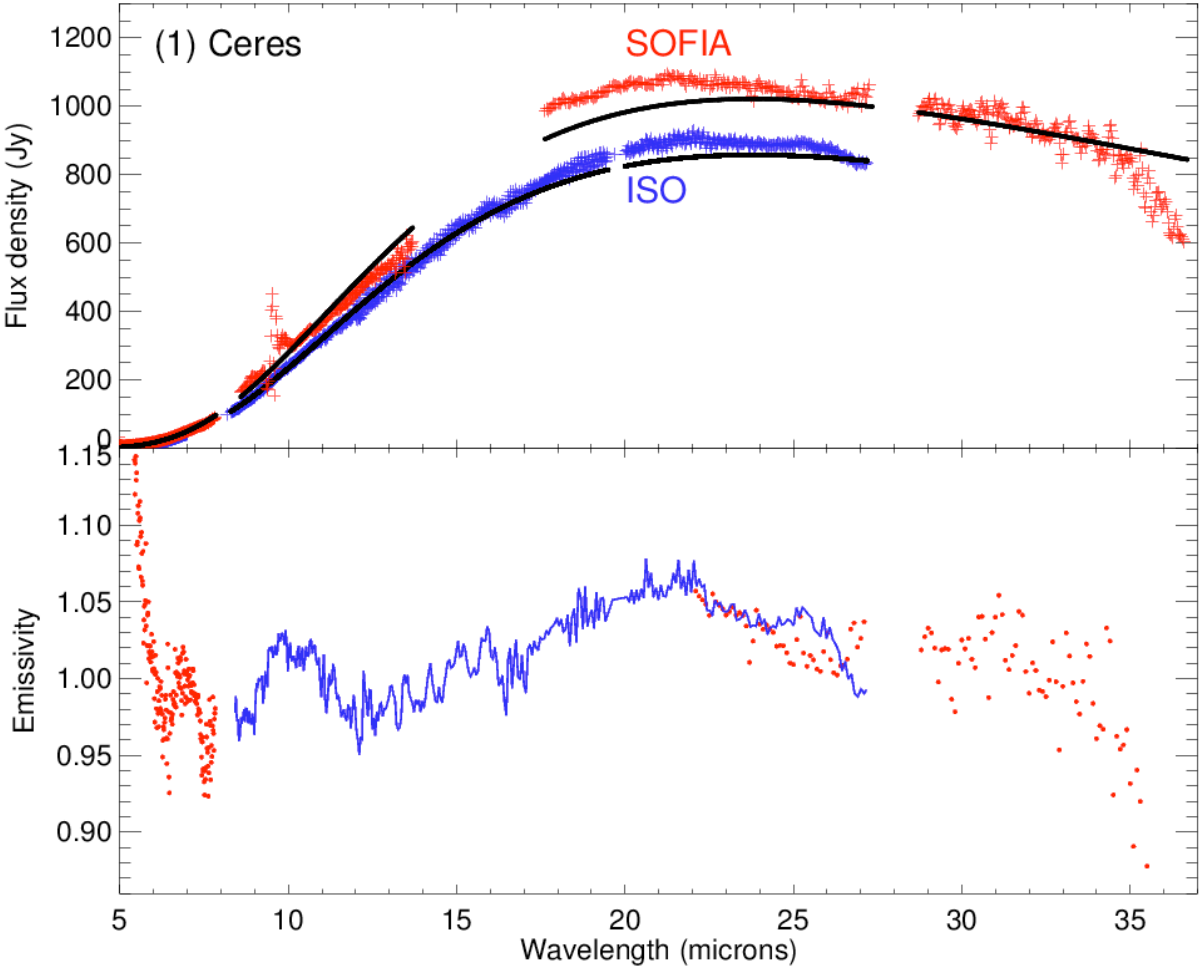

**Figure 1:** Upper panel: the SED of (1) Ceres obtained by ISO (blue curve) and SOFIA (red curve) with the best-fit thermal models (black curves). Lower panel: emissivity spectrum of (1) Ceres with the same color-coding as above.

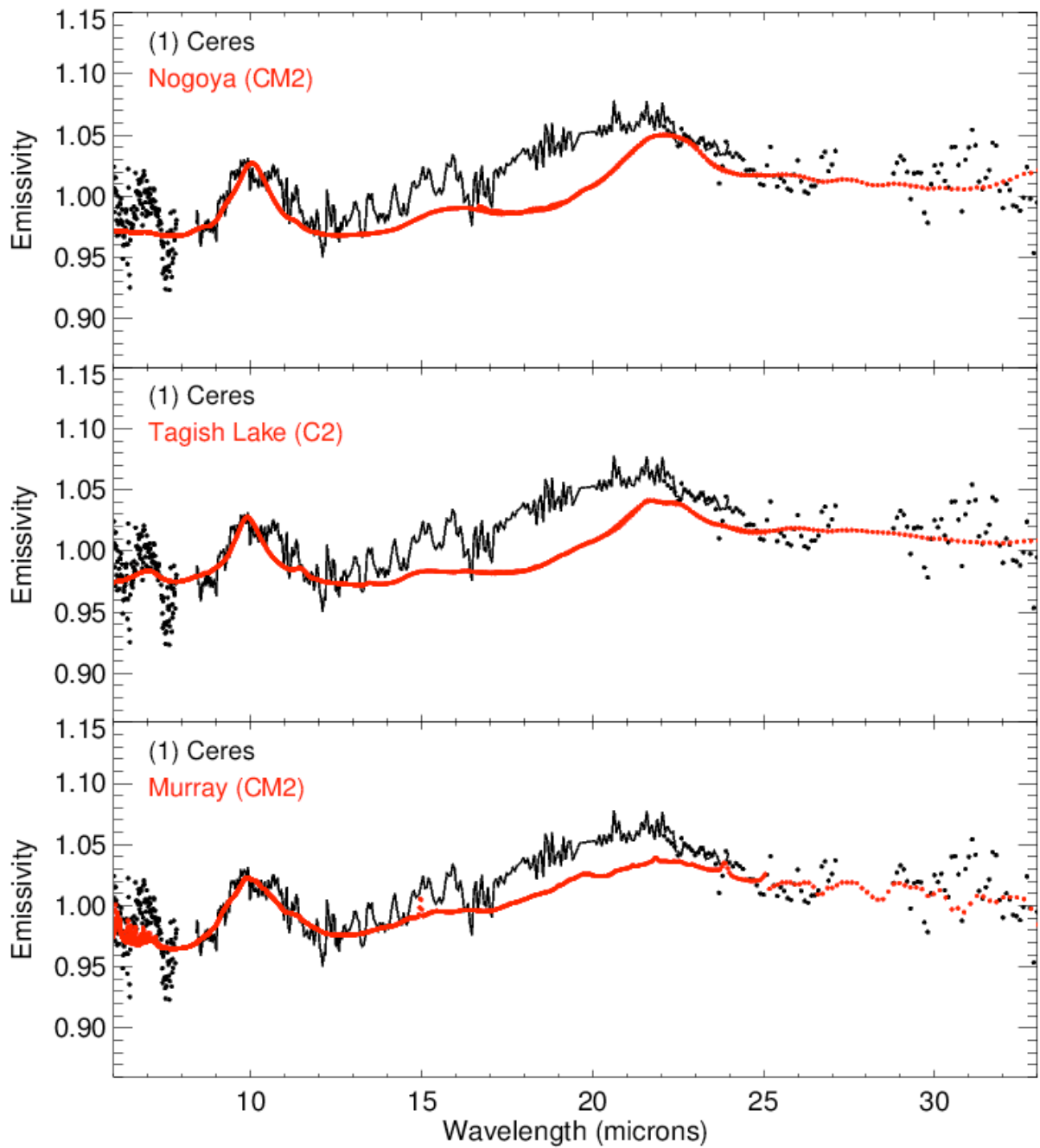

**Figure 2:** Comparison between the emissivity spectrum of (1) Ceres and the spectra of three carbonaceous chondrite meteorites. Whereas the two most prominent emissivity features in the Ceres spectrum (at ~10 μm and ~22 μm) are also seen in the meteorite spectra, their shape/profile appear quite different in both regions. Note that the mismatch in the 10-micron region is by far the more decisive with hydrated meteorites being single peaked whereas

Ceres is double peaked. This spectral difference suggests that an additional component is present at the surface of Ceres. Finally, a ~7 μm carbonate feature is observed both in the Ceres and meteorite spectra (especially in the Tagish Lake spectrum).

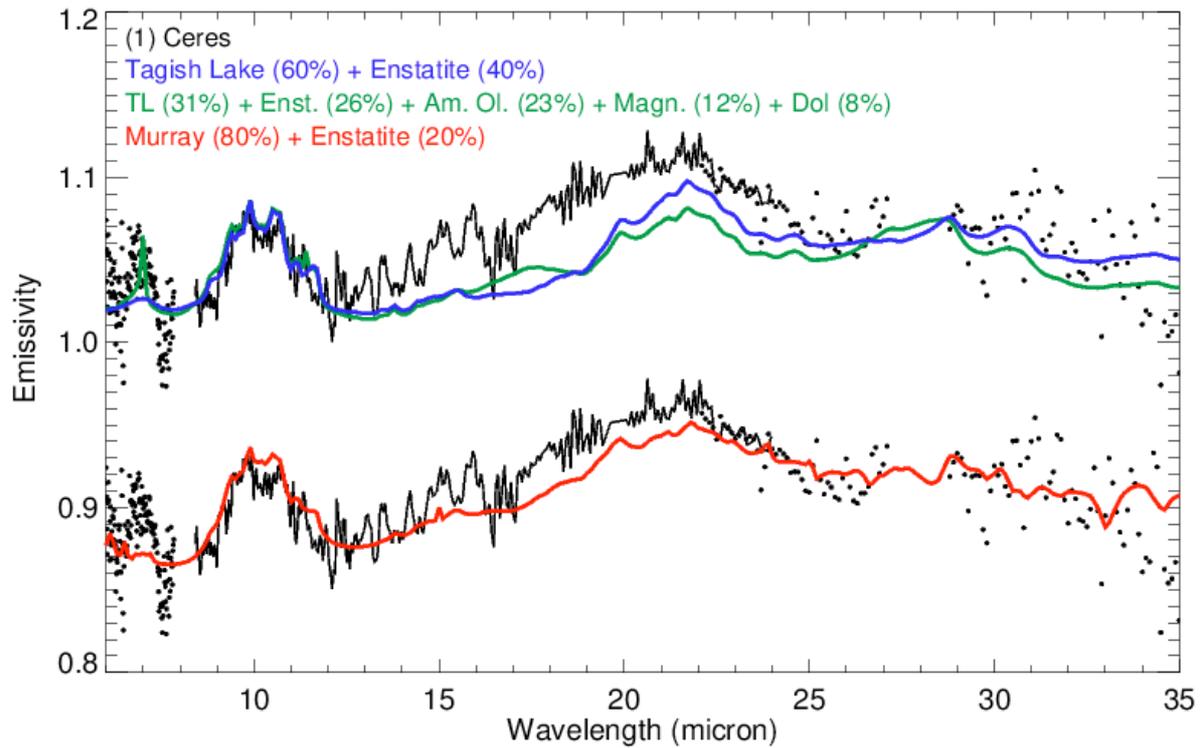

**Figure 3:** Comparison between the spectrum of (1) Ceres and two best-fit models. These models (blue and red curves) result from the linear combination of the emissivity spectra of enstatite and CI/CM chondrites (here the CM2 chondrite Murray and the CI/CM Tagish Lake). We also display an additional best-fit model (green curve) to highlight that additional components such as amorphous silicates of olivine-like composition and magnetite may also be present at the surface of Ceres (see text). Of prime importance, the best-fit requires at least 20% enstatite in the linear mixture.

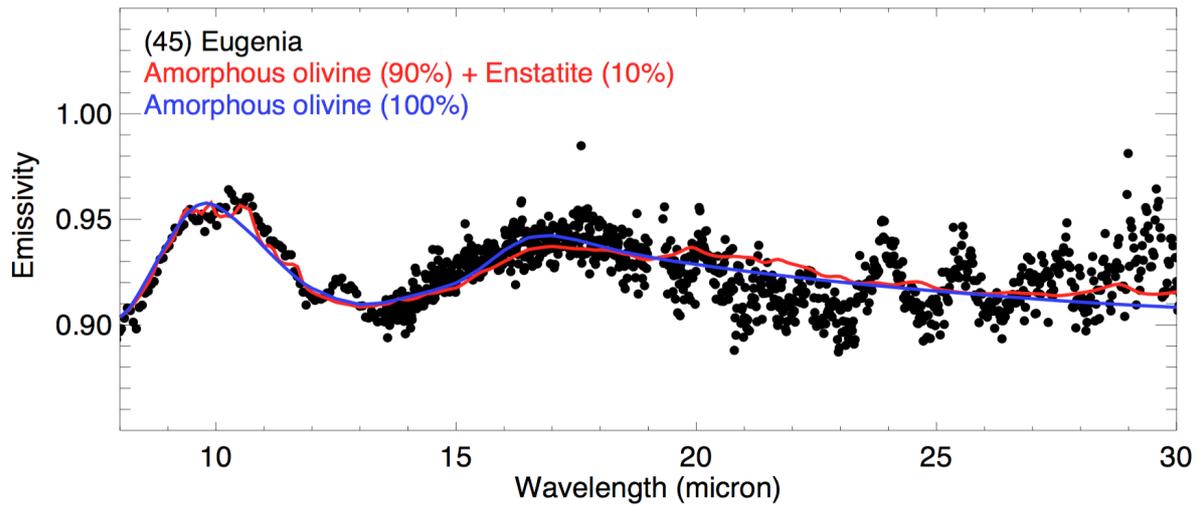

**Figure 4:** Comparison between the spectrum of (45) Eugenia and two best-fit models. Based on the model, which incorporates enstatite (red curve), it appears that both amorphous and crystalline components are present on the surface of Eugenia. Furthermore, the crystalline components are mostly pyroxene-rich.

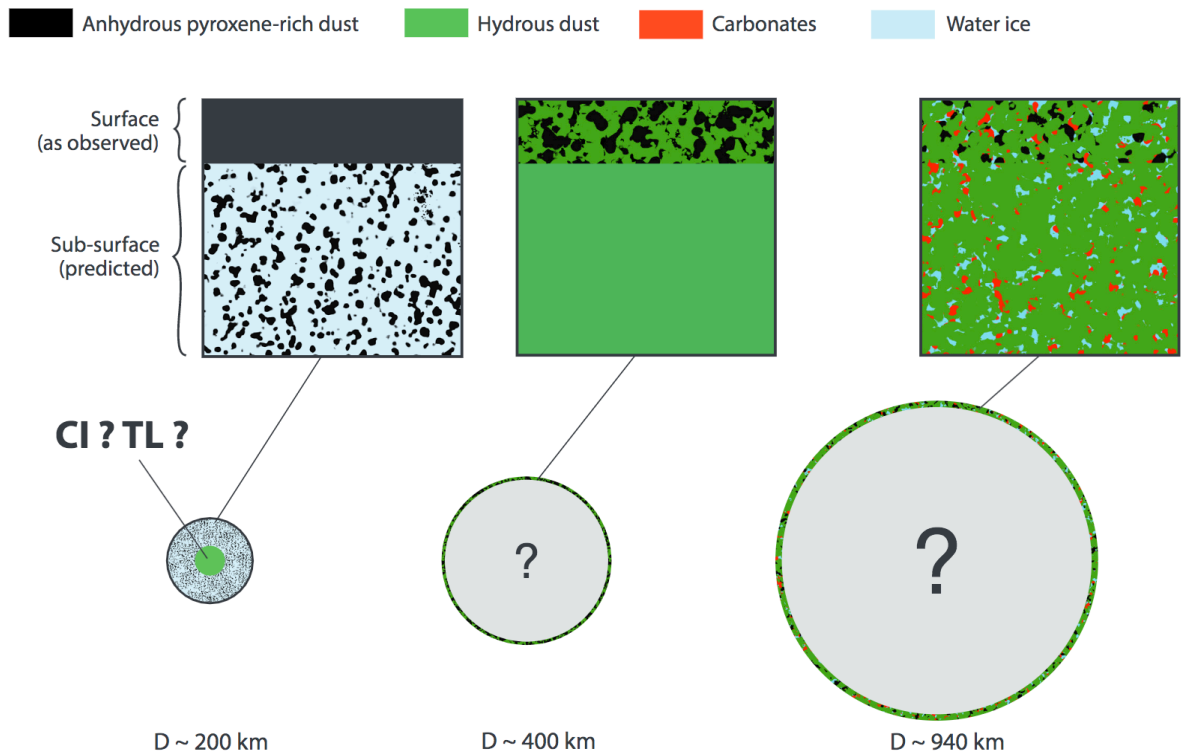

**Figure 5.** Illustration of the compositional diversity among C-type asteroid surfaces. Note that in the case of Ceres, we also highlight the presence of water ice within its crust as the latter component has been detected by the VIR spectrometer on board the Dawn mission (Combe et al. 2016a,b). In the case of Eugenia, we stress that the proposed internal structure is speculative at this stage and further work is required in order to i) confirm/infirm the presence of water ice in the interior of D<200km C-type asteroids and ii) test the hypothesis that CI chondrites, hydrated IDPs or the Tagish Lake (TL) meteorite could be samples of the cores of these bodies.

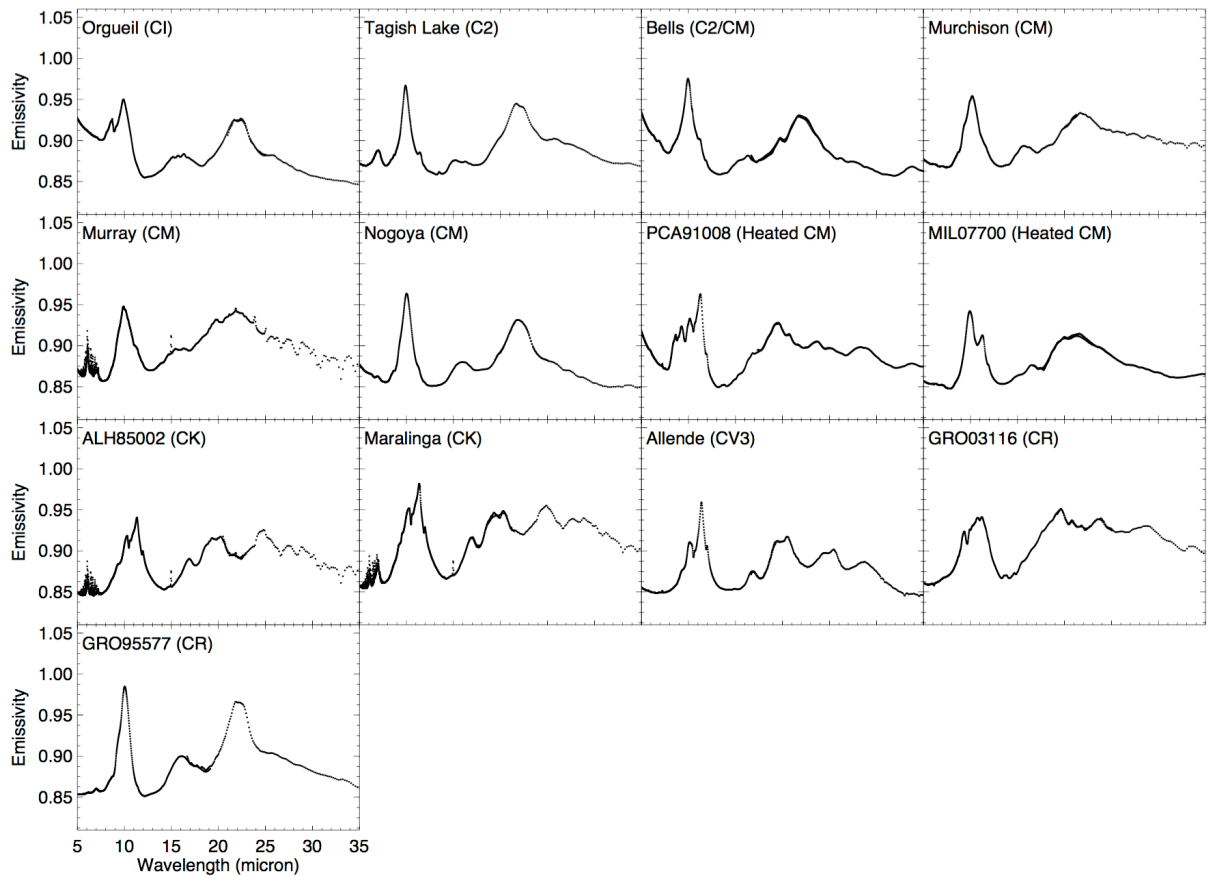

**Figure 6.** Mid-infrared spectral properties of carbonaceous chondrite meteorites collected over the 5-35 μm range. For each meteorite, a pellet was synthesized using 1.0 mg of meteorite dust dispersed in 300 mg of ultrapure KBr. The transmission Mid-IR (2-25 μm, resolution 2 cm$^{-1}$) spectra of the pellets were measured at Institut de Planétologie et d'Astrophysique de Grenoble (France; Beck et al. 2014) whereas the far-IR spectra were collected with the SMIS beamline (Dumas et al., 2006) of the synchrotron SOLEIL (France).

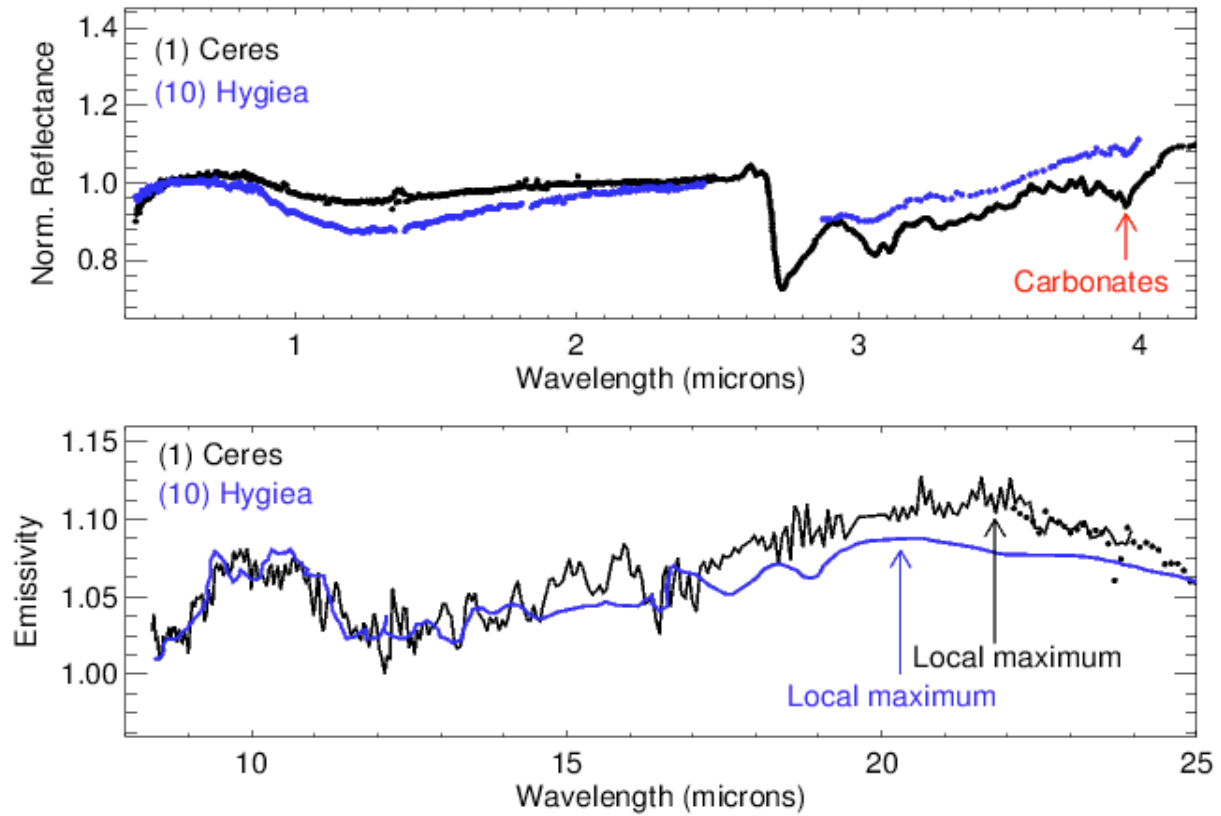

**Figure 7.** Comparison between the visible, near- and mid-infrared spectral properties of (1) Ceres and (10) Hygiea. The data were retrieved from http://smass.mit.edu/ and from the following papers: Takir & Emery (2012), Barucci et al. (2002), De Sanctis et al. (2015). Whereas the two asteroids possess very similar spectral properties, there are some noticeable differences between the two spectra: (i) the 3.06 μm band attributed to ammoniated phyllosilicates (De Sanctis et al. 2015) is much deeper in the spectrum of Ceres than in the Hygiea one, (ii) the 3.95 μm band due to the presence of carbonates is only unambiguously identified in the spectrum of Ceres, and (iii) the position of the 20 μm emissivity peak of Hygea is located at shorter wavelength with respect to that of Ceres.